\documentclass[12pt]{article}
\usepackage{bbm,latexsym,epsfig}
\newcommand{\lesssim}{ \ \mbox{\raisebox{-3pt}{$\stackrel%
{\displaystyle <}{\sim}$}} \ }

\newcommand{\ten}{\mathbf{10}}
\newcommand{\oht}{\mathbf{120}}
\newcommand{\hts}{\overline{\mathbf{126}}}
\newcommand{\B}{\mathcal{B}}
\newcommand{\mnu}{\mathcal{M}_\nu}
\newcommand{\deltasol}{\Delta m^2_\odot}
\newcommand{\deltaatm}{\Delta m^2_\mathrm{atm}}
\newcommand{\berr}{\!\begin{array}{l} \scriptstyle +}
\newcommand{\tr}{\\[-3mm] \scriptstyle -}
\newcommand{\eerr}{\end{array}}

\begin{document}

\title{\normalsize \hfill UWThPh-2006-29 \\[1cm]
\LARGE
A renormalizable $SO(10)$ GUT scenario \\ 
with spontaneous CP violation \\[8mm]}

\author{
Walter Grimus\thanks{E-mail: walter.grimus@univie.ac.at} \,
\normalsize and \large
\setcounter{footnote}{8}
Helmut K\"uhb\"ock\thanks{E-mail: helmut.kuehboeck@gmx.at}
 \\
\small Institut f\"ur Theoretische Physik, Universit\"at Wien \\
\small Boltzmanngasse 5, A--1090 Wien, Austria
\\*[4.6mm]}

\date{11 December 2006}

\maketitle

\begin{abstract} 
We consider fermion masses and mixings in a renormalizable SUSY 
$SO(10)$ GUT with Yukawa couplings of scalar fields in 
the representation $\ten \oplus \oht \oplus \hts$.
We investigate a scenario defined by the following assumptions: 
i) A single large scale in the theory, the GUT scale.
ii) Small neutrino masses generated by the type~I seesaw mechanism with
negligible type~II contributions.
iii) A suitable form of spontaneous CP breaking which induces 
hermitian mass matrices for all fermion mass terms of the Dirac type.
Our assumptions define 
an 18-parameter scenario for the
fermion mass matrices
for 18 experimentally known observables.
Performing a numerical analysis, we find excellent fits to all
observables in the case of both the normal and inverted neutrino mass
spectrum.  
\end{abstract}

\newpage

\section{Introduction}

The group $SO(10)$ is a favourite candidate for grand
unified theories (GUTs)~\cite{fritzsch} because its 16-dimensional irreducible
representation (irrep), the spinor representation, contains all chiral
fermions included in a Standard Model (SM) family plus an additional neutrino
SM gauge singlet. Moreover, such theories allow for 
type~I~\cite{seesaw} and type~II~\cite{typeII} seesaw mechanisms 
(see also~\cite{seesaw-general})
for the light neutrino masses. In the construction of $SO(10)$ theories, 
there are two options~\cite{so10-review}, 
either using low-dimensional scalar irreps but accepting
non-renormalizable terms in the Lagrangian, or one sticks to renormalizable
terms, then one has to accept high-dimensional scalar irreps 
according to~\cite{sakita,slansky}
\begin{equation}
\mathbf{16} \otimes \mathbf{16}
= \left( \mathbf{10} \oplus \mathbf{126} \right)_\mathrm{S}
\oplus \mathbf{120}_\mathrm{AS},
\label{tensor}
\end{equation}
where the subscripts ``S'' and ``AS'' denote, respectively,
the symmetric and antisymmetric parts of the tensor product.

In this paper, we deal with the second option. A special renormalizable model 
is the so-called ``minimal SUSY $SO(10)$ GUT'' (MSGUT)~\cite{MSGUT}, which 
uses, for the Yukawa couplings, one scalar in the $\mathbf{10}$
and one in the $\overline{\mathbf{126}}$ irrep in order 
to account for all fermion masses and mixings; it 
contains, in addition, one $\mathbf{210}$ and one $\mathbf{126}$ 
scalar irrep, in order to perform the suitable symmetry breakings.
This model 
has built-in the gauge-coupling unification
of the minimal SUSY extension of the Standard Model (MSSM).
Detailed studies of this minimal theory have been
performed~\cite{detailed,macesanu}, also with small 
effects of the 120-plet~\cite{small120}.
Though the MSGUT works very well in the fermion
sector, there is a tension between the scale of the light
neutrino masses and the GUT scale. The reason is that the
natural order of the neutrino masses in GUTs is 
$v^2/M_\mathrm{GUT} \sim 1.5 \times 10^{-3}$ eV, 
where we have used $v\sim 174$ GeV for the electroweak scale and a GUT scale
of $M_\mathrm{GUT} \sim 2 \times 10^{16}$ GeV. 
This neutrino mass scale is too low because 
$\sqrt{\deltaatm} \sim 0.05$ eV 
where $\deltaatm$ is the atmospheric neutrino mass-squared
difference---for reviews on the status of neutrino masses and mixing 
see~\cite{nu-review,schwetz1}.
Studies of the 
heavy scalar states~\cite{heavyscalarspectrum}, together with studies of the
fermion mass spectrum, have shown that the MSGUT
is too constrained~\cite{au05}, and the tension between the scale of light
neutrino masses and the GUT scale cannot be overcome: if one has a
good fit of the fermion masses, which requires a seesaw scale below
$M_\mathrm{GUT}$,  then the gauge coupling unification of
the MSSM~\cite{garg,schwetz} is spoiled. 

A natural step for supplying additional degrees of freedom to the MSGUT is to
add the 120-plet of scalars~\cite{aulakh06} 
which appears anyway in Eq.~(\ref{tensor})---for early works in this direction
see~\cite{oshimo,yang,dutta1,dutta2}.\footnote{We stress 
that $\ten \oplus \oht$
alone does not give a good fit in the charged fermion
sector~\cite{LKG}.} The disadvantage is that this step adds a considerable
number of parameters and reduces the predictability of the theory.
Adding the $\oht$ leads to a resurgence of the type~I seesaw
mechanism~\cite{macesanu}, as a consequence of the collapse of the
seesaw scale with the GUT scale, because type~I
seesaw allows to enhance the neutrino masses 
through small Yukawa couplings of the 
$\hts$~\cite{aulakh06,GK,aulakh-120}; without the $\oht$, i.e.
in the MSGUT, this process leads to the contradictions mentioned above.

The $\oht$ has electrically neutral components only
in its four doublets with respect to the SM gauge group.
These contribute to the Higgs doublets $H_d$, $H_u$ of the MSSM, 
which are assumed to be the only
light scalar degrees of freedom and the only ones which 
acquire VEVs at the electroweak scale. 
Thus the MSGUT enlarged by the $\oht$ 
inherits, from the MSGUT, the scalar fields responsible for
spontaneous symmetry breaking above the electroweak scale.
In~\cite{GK} we took this 
into account by explicitly making the identification
\begin{equation}\label{onescale}
w_R = M_\mathrm{GUT} = 2 \times 10^{16}\: \mbox{GeV},
\end{equation}
where $w_R$, which defines the seesaw scale, 
is the vacuum expectation value (VEV) of 
$(\mathbf{10}, \mathbf{1}, \mathbf{3}) \in \overline{\mathbf{126}}$,
with the usual notation for multiplets of the Pati--Salam
subgroup~\cite{PS} 
$SU(4)_c \times SU(2)_L \times SU(2)_R$
of $SO(10)$.
Furthermore, in that paper we reduced the number of parameters in the fermion
mass matrices by assuming a horizontal $\mathbbm{Z}_2$ symmetry and 
spontaneous CP violation, i.e. real Yukawa
couplings, with CP violation stemming from the phases of the VEVs.
We showed numerically that this scenario can excellently reproduce
the known fermion masses and mixings.

Recently, the role of spontaneous CP violation has been upgraded. 
A ``New MSGUT'' (NMSGUT) was proposed~\cite{newMSGUT},
defined by extending the MSGUT by the $\oht$ and spontaneous CP
violation. It was shown that the requirement of spontaneous CP
violation not only has the virtue of reducing the number of parameters
of the theory but it has an important impact, via  threshold effects,
on the unification scale as well; it tends to raise the unification
scale and with it the masses of all heavy multiplets, thereby
suppressing baryon decay. 

In the present paper we retain Eq.~(\ref{onescale}) as a reference point.
We do not employ any horizontal symmetry but we
again motivate real Yukawa coupling matrices by spontaneous CP
violation. However, we assume that it is of a very specific kind: 
CP is solely violated by imaginary VEVs of the $\oht$; 
the VEVs of the $\ten$ and $\hts$ are assumed to be real. 
In this way, the mass matrices of the down-quarks, up-quarks,
charged leptons and the neutrino Dirac-mass matrix are hermitian. This
scenario was originally proposed in~\cite{dutta1}, its compatibility with
sufficiently slow proton decay shown in~\cite{dutta2}. 
However, in~\cite{dutta1,dutta2} it was assumed that the type~II seesaw
mechanism is dominating. 
Since this is incompatible with having only one large scale, 
we have in the present paper type~I dominance and neglect possible small
contributions of type~II, suppressed by $v^2/M_\mathrm{GUT}$. 
Our scenario gives an excellent fit to all known fermion masses, mixings and
the CKM phase $\delta_\mathrm{CKM}$, as good as the one in~\cite{GK}, though
it is of a rather different type. This shows that 
the fermion data do not fix the
enlarged MSGUT in a unique way and there is considerable freedom in reducing
the number of parameters in this theory.

The paper is organized as follows.
In Section~\ref{CPviolation} we discuss CP-invariant Yukawa couplings and lay
out our scenario. The method and results of our numerical analysis
are discussed in Section~\ref{numerical}. 
In Section~\ref{concl} we present the conclusions.
Appendix~\ref{so10} contains a small collection of formulas for the $SO(10)$ 
spinor representation, which is helpful for Section~\ref{CPviolation}.

\section{An $SO(10)$ scenario motivated by spontaneous 
CP violation}
\label{CPviolation}

Let us a define a transformation
\begin{equation}\label{CPgauge}
\mbox{CP:} \quad 
\begin{array}{ccl}
\psi_L (x) & \to & i\,C \psi_L^* (\hat x), \\[1mm]
W^{pq}_\mu(x) & \to & \varepsilon(\mu) \eta_{pq} W^{pq}_\mu(\hat x),
\end{array}
\end{equation}
where $\psi_L$ is a fermionic 16-plet, 
$C$ is the charge-conjugation matrix, the 45 gauge fields are
denoted by $W^{pq}_\mu$ ($p < q$),  
$\varepsilon(\mu) = 1$ for $\mu = 0$ and $-1$ for $\mu = 1,\,2,\,3$, 
$\hat x = ( x^0, -\vec x)$, and the $\eta_{pq}$ are signs. No
summation is implied in Eq.~(\ref{CPgauge}).
The 45 (hermitian) 
generators of the gauge group $SO(10)$ in the fermionic
$\mathbf{16}$ are given by
\begin{equation}\label{generators}
\frac{i}{2} \sigma_{pq} = \frac{i}{2}\, \Gamma_p \Gamma_q \quad 
(1 \leq p < q \leq 10).
\end{equation}
For a representation of the operators $\Gamma_p$ ($p = 1, \ldots, 10)$
of the Clifford algebra and useful formulas
concerning the spinor irrep $\mathbf{16}$ see Appendix~\ref{so10}.
One can easily check that the gauge
interaction of the fermionic 16-plet is invariant under the
transformation~(\ref{CPgauge}) if~\cite{slansky,smolyakov,grimus}
\begin{equation}
- \sigma_{pq}^T\,\eta_{pq} = \sigma_{pq}.
\end{equation}
Since
\begin{equation}
\Gamma_p^T = \xi_p \Gamma_p \quad \mbox{with} \quad \xi_p =
(-1)^{1+p},
\end{equation}
one finds
\begin{equation}
\eta_{pq} = \xi_p \xi_q = (-1)^{p+q}.
\end{equation}
Denoting the generators of the Lie algebra $so(10)$ by $M_{pq}$,
we mention that 
\begin{equation}\label{auto-cp}
M_{pq} \to \eta_{pq}\, M_{pq} = S M_{pq} S
\quad \mbox{with} \quad 
S = \mbox{diag}\,
(1,-1,1,-1,1,-1,1,-1,1,-1)
\end{equation}
is the automorphism associated with root reflection, which is
the canonical automorphism associated with CP. Such an automorphism
exists for all compact Lie groups and is the reason why any gauge Lagrangian,
whether for fermions or scalars, is
CP-invariant~\cite{slansky,smolyakov,grimus}. 

Now we transfer the CP transformation to the Yukawa couplings
given by the Lagrangian
\begin{eqnarray}
\mathcal{L}_Y & = & \frac{1}{2} \left(  
H_{ab}\, \psi_{aL}^T C^{-1} \B\, \Gamma_p H_p \psi_{bL} + 
G_{ab}\, \psi_{aL}^T C^{-1} \B\, \Gamma_p \Gamma_q \Gamma_r 
D_{pqr} \psi_{bL} + \right. \nonumber \\ && \left. 
\hphantom{xxi}
F_{ab}\, \psi_{aL}^T C^{-1} \B\, \Gamma_p \Gamma_q \Gamma_r \Gamma_s \Gamma_t
{\bar\Delta}_{pqrst} \psi_{bL} \right) + \mbox{H.c.}
\label{YukawaL}
\end{eqnarray}
The indices $a$, $b$ denote the family indices,
$p,\, q,\, r,\, s,\,t = 1, \ldots, 10$ are $SO(10)$ indices,
and $C$ is the charge-conjugation matrix.
Summation over family and $SO(10)$ indices is implied in
Eq.~(\ref{YukawaL}). 
The matrix $\B$, which ensures $SO(10)$ invariance, 
is defined in Eq.~(\ref{B}).
The Yukawa coupling matrices have the properties
\begin{equation}\label{Yprop}
H_{ab} = H_{ba}, \quad G_{ab} = -G_{ba}, \quad 
F_{ab} = F_{ba}.
\end{equation}
We define CP transformations (no summations implied)
\begin{eqnarray}
H_p (x) & \to & \xi_p H_p^*(\hat x),
\label{CP10} \\
D_{pqr} (x) & \to & \xi_p \xi_q \xi_r D_{pqr}^*(\hat x), 
\label{CP120} \\
\bar \Delta_{pqrst} (x) & \to & \xi_p \xi_q \xi_r \xi_s \xi_t 
\bar \Delta_{pqrst}^*(\hat x),
\label{CP126}
\end{eqnarray}
for the scalar fields of the irreps $\ten$, $\oht$ and $\hts$,
respectively; the latter two are totally antisymmetric tensor fields,
$\bar \Delta$ is self-dual in addition.

Now we require invariance of the Lagrangian~(\ref{YukawaL}) under the
CP transformation given by Eqs.~(\ref{CPgauge}), (\ref{CP10}),
(\ref{CP120}) and (\ref{CP126}). As an example we take the $\ten$ and obtain
\begin{equation}\label{CPH}
H_{ab}\, \psi_{aL}^T \B C^{-1} \Gamma_p H_p \psi_{bL} 
\stackrel{\mathrm{CP}}{\to} 
-H_{ab}\, \psi_{aL}^\dagger \B C \Gamma_p \xi_p H_p^* \psi_{bL}^* = 
\left( H_{ba} \psi_{bL}^T \B C^{-1} \Gamma_p H_p \psi_{aL} \right)^\dagger.
\end{equation}
The equality sign on the right-hand side of the arrow defines the condition
of CP invariance: the CP-transformed Yukawa Lagrangian must be
identical with its hermitian conjugate. 
Evaluating Eq.~(\ref{CPH}) with the help of Eq.~(\ref{B}),
we find a hermitian Yukawa coupling matrix.
Performing an analogous computation for the $\oht$ and $\hts$ we arrive
at the conclusion that the CP transformation requires 
\begin{equation}
H_{ab} = H_{ba}^*, \quad 
G_{ab} = -G_{ba}^*, \quad 
F_{ab} = F_{ba}^*.
\end{equation}
Together with Eq.~(\ref{Yprop}) this means that \emph{all} Yukawa
coupling matrices are real.

In order to obtain a non-trivial CKM phase $\delta_\mathrm{CKM}$ it is
necessary to break CP invariance. The scenario we envisage was
originally proposed in~\cite{dutta1}. In the context discussed here,
we assume that 
\begin{itemize}
\item
the VEVs of the $\ten$ and $\hts$ are \emph{real},
\item
CP is spontaneously broken by the VEVs of the $\oht$,
\item
this breaking is maximal, i.e., 
the VEVs of the $\oht$ are imaginary.
\end{itemize}
Thus, the mass matrices of 
the charged fermions 
and the neutrino Dirac-mass matrix
are given, respectively, by
\begin{eqnarray}
M_d    & = & k_d\, H + i\kappa_d\,    G +   v_d\, F, 
\label{md} \\
M_u    & = & k_u\, H + i\kappa_u\,    G +   v_u\, F, 
\label{mu} \\
M_\ell & = & k_d\, H + i\kappa_\ell\, G - 3 v_d\, F, 
\label{ml} \\
M_D    & = & k_u\, H + i\kappa_D\,    G - 3 v_u\, F,
\label{mD}
\end{eqnarray}
with
\begin{equation}
k_{d,u},\,  \kappa_{d,\ell,u,D},\, v_{d,u} \in \mathbbm{R}, \quad 
H = H^* = H^T, \; G = G^* = -G^T, \; F = F^* = F^T.
\end{equation}
The mass matrices~(\ref{md})--(\ref{mD}) are hermitian.
The light neutrino mass matrix is given by
\begin{equation}\label{mnu}
\mnu = M_L - M_D^T M_R^{-1} M_D
\quad \mbox{with} \quad
M_L = w_L\, F, \quad
M_R = w_R\, F, 
\end{equation}
with scalar triplet VEVs $w_L$ and $w_R$.
The mass Lagrangian of the ``light'' fermions reads 
\begin{equation}
\mathcal{L}_M  = 
- \bar d_R M_d\, d_L - \bar u_R M_u\, u_L - \bar \ell_R M_\ell\, \ell_L +
\frac{1}{2} \nu_L^T C^{-1}\mnu \nu_L + \mbox{H.c.}
\end{equation}

We finish this section with some remarks.
$SO(10)$ models have included the so-called D-parity~\cite{Cha84},
which is a specific involutory $SO(10)$ transformation which
uses the branching rule
\begin{equation}
\mathbf{16} = 
(\mathbf{4}, \mathbf{2}, \mathbf{1}) \oplus
(\bar\mathbf{4}, \mathbf{1}, \mathbf{2})
\end{equation}
under the Pati--Salam group~\cite{PS} and exchanges the two 
Pati--Salam irreps.
One can combine CP with D-parity and interpret
such a transformation 
as parity in the usual sense~\cite{grimus}. 
Requiring invariance of the theory
under this parity, gives the same restrictions on the Yukawa coupling
matrices as CP alone, since the theory is invariant under D-parity anyway.

In the CP transformation of the $\oht$, Eq.~(\ref{CP120}), we could 
put a minus sign. Then the Yukawa coupling matrix would be hermitian
and antisymmetric or, equivalently, antisymmetric and
imaginary.\footnote{This is the choice in~\cite{dutta1}.}
In that case, real VEVs of the $\oht$ break CP maximally.

\section{The numerical analysis}
\label{numerical}

\begin{table}[t]
\begin{center}
\renewcommand{\arraystretch}{1.2}
\begin{tabular}{cc}
\begin{tabular}[t]{|c|c|} \hline
\multicolumn{2}{|c|}{Quarks} \\ \hline\hline
$m_d$ &  
$1.03 \pm 0.41$ \\ \hline
$m_s$ & $19.6 \pm 5.2$ \\ \hline
$m_b$ & $1063.6 \berr 141.4 \tr 086.5 \eerr$ \\ \hline
$m_u$ & $0.45 \pm 0.15$ \\ \hline
$m_c$ & $210.3273 \berr 19.0036 \tr 21.2264 \eerr$ \\ \hline
$m_t$ & $82433.3 \berr 30267.6 \tr 14768.6 \eerr$ \\ \hline
$s_{12}$ & $0.2243 \pm 0.0016$ \\ \hline
$s_{23}$ & $0.0351 \pm 0.0013$ \\ \hline
$s_{13}$ & $0.0032 \pm 0.0005$ \\ \hline
$\delta_\mathrm{CKM}$ & $60^\circ \pm 14^\circ$ \\ \hline
\end{tabular}
&
\begin{tabular}[t]{|c|c|} \hline
\multicolumn{2}{|c|}{Leptons} \\ \hline\hline
$m_e$       &  
$0.3585 \berr 0.0003 \tr 0.0003 \eerr$ \\ \hline
$m_\mu$     & 
$75.6715 \berr 0.0578 \tr 0.0501 \eerr$ \\ \hline
$m_\tau$    & 
$1292.2 \berr 1.3 \tr 1.2 \eerr$ \\ \hline
$\deltasol$ & $(7.9 \pm 0.3) \times 10^{-5}$ \\ \hline
$\deltaatm$ & $\Big(2.50 \berr 0.20 \tr 0.25 \eerr 
\Big) \times 10^{-3}$ \\ \hline
$s_{12}^2$  & $0.31 \pm 0.025$ \\ \hline
$s_{23}^2$  & $0.50 \pm 0.065$ \\ \hline
$s_{13}^2$  & $< 0.0155$ \\ \hline
\end{tabular}
\end{tabular}
\end{center}
\caption{Input data at the GUT scale for $M_\mathrm{GUT} = 2 \times
  10^{16}$ GeV and $\tan \beta = 10$. The charged-fermion masses are
  taken from~\cite{das}, except for the values of 
  $m_d$, $m_s$ and $m_u$; these were obtained by taking their
  low-energy values from~\cite{RPP} and scaling them to
  $M_\mathrm{GUT}$. As for $\deltaatm$, we use the value obtained
  in~\cite{schwetz1}. 
  We have copied the remaining input from Table~I in~\cite{schwetz}.
  Charged-fermion masses are in units of MeV,
  neutrino mass-squared differences in eV$^2$. We have used the
  abbreviations $s_{12} \equiv \sin \theta_{12}$, etc. 
  The angles in the left table refer to the CKM matrix, in the right
  table to the PMNS matrix.\label{input}}
\end{table}
As argued in the introduction, with only one large scale, the GUT
scale, in the theory, we can neglect the type~II seesaw contribution
in Eq.~(\ref{mnu}).\footnote{A quantitative justification will be
  given in this section.} Then a possible phase of $w_R$ is irrelevant.
With
\begin{equation}\label{M'}
H' \equiv k_d H, \quad G' \equiv \kappa_d G, \quad
F' \equiv v_d F,
\end{equation}
we rewrite the mass matrices as 
\begin{eqnarray}
M_d    & = & H' + iG' +  F', 
\label{Md} \\
M_u    & = & r_H H' + ir_u\, G' + r_F F', 
\label{Mu} \\
M_\ell & = & H' + ir_\ell\,G' - 3\, F', 
\label{Ml} \\
M_D    & = & r_H H' + ir_D\,G' - 3\, r_F F', 
\label{MD} \\
\mnu   & = & r_R\, M_D^T {F'}^{-1} M_D.
\label{Mnu} 
\end{eqnarray}
Without loss of generality we assume $H'$ to be diagonal. Then all
redundant parameters are removed and we arrive at 12 real parameters
in $H'$, $G'$, $F'$ and 
six real ratios of VEVs. Thus our scenario has 18
independent parameters for 18 observables: 
nine charged-fermion masses, three mixing angles and 
the CP phase $\delta_\mathrm{CKM}$ in the CKM matrix, the atmospheric
and solar 
neutrino mass-squared differences $\deltaatm$ and $\deltasol$, and three
lepton mixing angles.

Eqs.~(\ref{Md})--(\ref{Mnu}) are amenable to a numerical analysis,
which will, in particular, yield values for $r_F$ and $r_R$. 
If we fix the triplet VEV $w_R$, e.g. by identifying it with the GUT
scale---see Eq.~(\ref{onescale}), this analysis will also yield
definite values for $v_d$ and $v_u$ because
\begin{equation}\label{vud}
v_d = r_R w_R, \quad v_u = r_F r_R w_R.
\end{equation}
A reasonable condition on these VEVs is given 
by~\cite{GK} 
\begin{equation}\label{test}
v_d^2 + v_u^2 =
\left( r_R w_R \right)^2 \left( 1 + r_F^2 \right) < v^2 
\quad \mbox{with} \quad v = 174\: \mathrm{GeV}.
\end{equation}
This inequality certainly holds at the electroweak scale. Assuming
that it holds approximately at the GUT scale as well, we will subject
our fit results to this consistency check.

To find a numerical solution for the parameters in
Eqs.~(\ref{Md})--(\ref{Mnu}), we build as usual~\cite{schwetz,LKG,GK} a
$\chi^2$-function for the 18 observables, 
\begin{equation}\label{chi2}
\chi^2(P) = \sum_{i=1}^{18} \left( \frac{f_i(P) - \bar y_i}{\delta y_i}
\right)^2,
\end{equation}
whose input values $\bar y_i \pm \delta y_i$
are given in Table~\ref{input}; these values refer to an MSSM
parameter $\tan \beta = 10$. The letter $P$ symbolizes the set of 18
parameters, i.e. the Yukawa couplings and ratios of VEVs.
The functions $f_i(P)$ express our theoretical predictions, as
functions of the parameter set $P$, for the
observables, obtained from Eqs.~(\ref{Md})--(\ref{Mnu}).
As convention for the quark and lepton mixing matrix we use that of
the Review of Particle Properties~\cite{RPP}. 
The $\chi^2$-function is minimized
analytically with respect to $r_R$. In this way we obtain a
$\chi^2$-function of the remaining 17 parameters, which is minimized
numerically by employing the downhill simplex method~\cite{downhill}.

In the following we will also consider $\chi^2$ functions where a specific
quantity is pinned down to a given value---for previous uses of such
a method see e.g.~\cite{schwetz,GK}. 
If we want to pin down a quantity $\omega(P)$, which is independent of
the 18 observables, to a value $\bar \omega$, we add 
$( \omega(P) - \bar \omega )^2/(0.01 \bar \omega)^2$ in Eq.~(\ref{chi2})
and minimize the thus obtained $\chi^2_\omega$. 
If $\omega$ coincides with one of the observables $y_k$, the term above is
added but at the same time the term 
$(f_k(P) - \bar y_k)^2/(\delta y_k)^2$ has to be removed from the
$\chi^2$ of Eq.~(\ref{chi2}).
In that way, we can study the sensitivity of our
scenario to a variation of a quantity $\omega$.

\begin{figure}[t]
\begin{center}
\epsfig{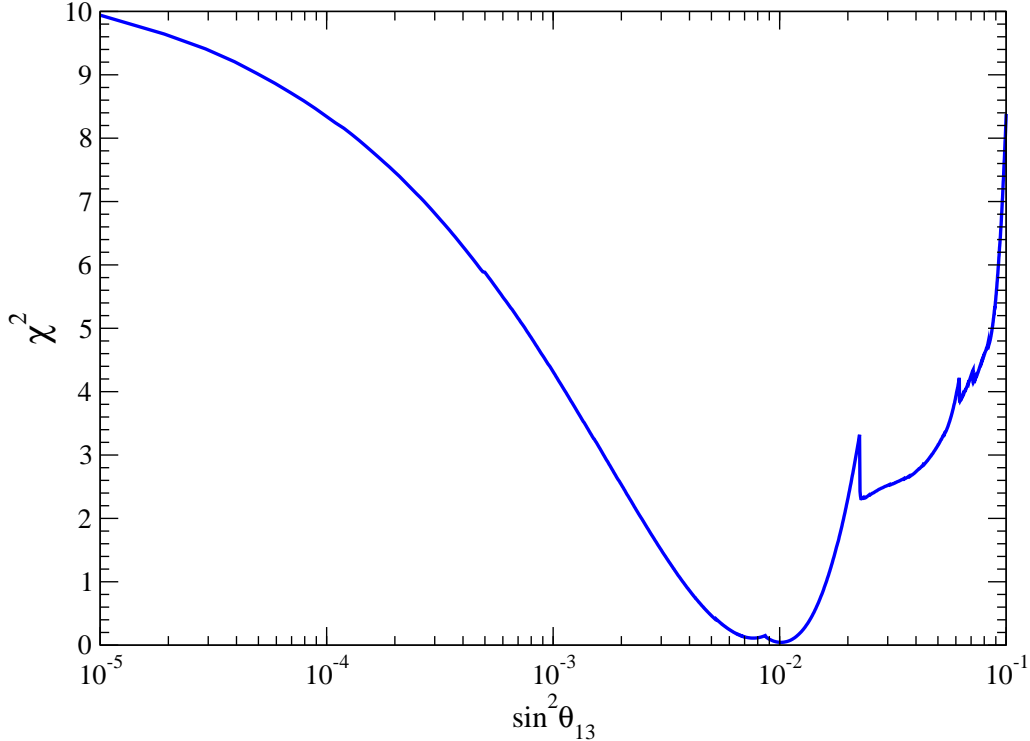}
\end{center}
\caption{$\chi^2$ as a function of the leptonic
  $\sin^2 \theta_{13}$ for the normal neutrino mass spectrum. \label{fig13}}
\end{figure}
\begin{figure}[t]
\begin{center}
\epsfig{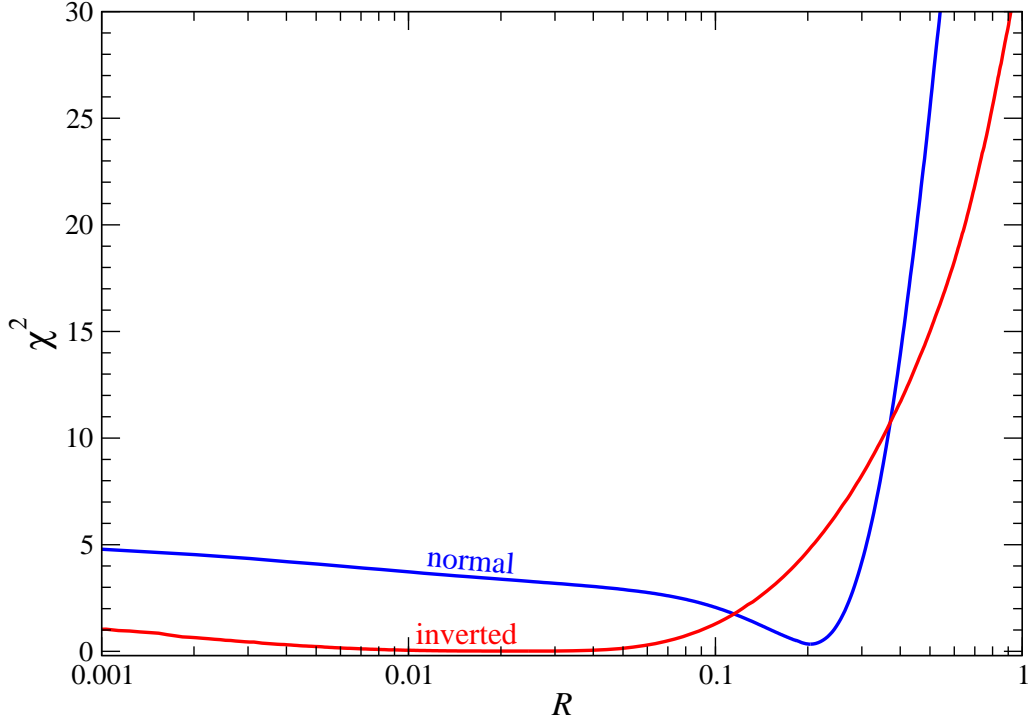}
\end{center}
\caption{$\chi^2$ as a function of 
  $R = m_\mathrm{min}/\sqrt{\deltasol}$, where $m_\mathrm{min} =
  m_1$ for the normal and $m_3$ for the inverted neutrino mass
  spectrum. \label{figR}}
\end{figure}
\begin{table}[t]
\begin{center}
\renewcommand{\arraystretch}{1.5}
\begin{tabular}{r|c|c|c|c|c|c}
& $r_H$ & $r_F$ & $r_u$ & $r_\ell$ & $r_D$ & $r_R$ \\ \hline
Normal: &
$-71.5516$ &
$7.16038$ &
$-2.76118$ &
$6.57185$ &
$5842.373$ &
$1.82618 \times 10^{-16}$
\\[-2mm]
Inverted: &
$82.0042$ &
$190.194$ &
$6.60808$ &
$-7.22108$ &
$-43303.97$ &
$2.22386 \times 10^{-18}$
\end{tabular}
\end{center}
\caption{The values of the VEV ratios appearing in the mass
  matrices~(\ref{Md})--(\ref{Mnu}) obtained in the fits with 
  normal and inverted neutrino mass ordering. \label{ratios}}
\end{table}
\paragraph{A fit in the case of normal neutrino mass ordering:}
We search for a solution in the case of the normal ordering 
$m_1 < m_2 < m_3$ of the neutrino masses 
($\deltasol = m_2^2 - m_1^2$, $\deltaatm = m_3^2 - m_1^2$).
In that case we find an excellent fit with the following properties:
\begin{equation}\label{fit-normal}
\begin{array}{l}
\chi^2 = 0.33, \quad
\sqrt{v_d^2 + v_u^2} = 26.4\: \mbox{GeV}, \\[1mm]
m_1 = 1.81\times 10^{-3} \: \mbox{eV}, \quad
m_2 = 0.907 \times 10^{-2} \: \mbox{eV}, \quad
m_3 = 5.006 \times 10^{-2} \: \mbox{eV}. 
\end{array}
\end{equation}
In the second relation in this equation we have used $w_R =
M_\mathrm{GUT}$---see Eq.~(\ref{onescale}).
The corresponding values of the matrix elements of $H'$, $G'$, $F'$
are given by
\begin{eqnarray}
H' & = & \left( \begin{array}{ccc}
0.198268 & 0 & 0 \\ 
0 & -0.896317 & 0 \\ 
0 & 0 & 1150.786
\end{array} \right), \nonumber \\
G' & = & \left( \begin{array}{ccc}
0 & 2.68402 & 1.67616 \\ 
-2.68402 & 0 & -0.580726 \\ 
-1.67616 & 0.580726 & 0 
\end{array} \right),
\label{solution-normal} \\
F' & = & \left( \begin{array}{ccc}
3.23879 & 5.69390 & -10.0743 \\ 
5.69390 & 19.1109 & -33.5646 \\ 
-10.0743 & -33.5646 & -44.3762
\end{array} \right), \nonumber
\end{eqnarray}
where all numerical values are in units of MeV.
The fit values of the VEV ratios are listed in Table~\ref{ratios}.
The $\chi^2$ of the fit practically comes only from two observables: 
the pull of $\sin^2 \theta_{13}$ (leptonic mixing angle) is 0.45  and
the pull of $m_b$ is 0.31. 

We can ask the question if our scenario makes some predictions. 
For the best fit we find $\delta_\mathrm{PMNS} = -2.0^\circ$. 
However, this small value misleading, 
because pinning $\delta_\mathrm{PMNS}$ in $\chi^2$ shows that in the large
range $-60^\circ \lesssim \delta_\mathrm{PMNS} \lesssim 60^\circ$ the
fit is still very good, 
with $\chi^2 \lesssim 1$. The worst $\chi^2$ is about 15 and occurs
around $\delta_\mathrm{PMNS} \sim 175^\circ$, where for instance $m_b$ is not
well reproduced and the leptonic $\sin^2 \theta_{13}$ becomes too large.
As for $\sin^2 \theta_{13}$, Fig.~\ref{fig13} shows that the preferred value
is about 0.01. However, we cannot consider this as a prediction since in a
wide range around this value the $\chi^2$ is still acceptable. Only at very 
small $\sin^2 \theta_{13}$ the fit becomes bad, mainly because of $m_d$, $m_b$
and the atmospheric mixing angle $\theta_{23}$.
The quantity $R = m_1/\sqrt{\deltasol}$ measures how hierarchical a normal
neutrino mass spectrum is. The $\chi^2$ as a function of $R$ is depicted in
Fig.~\ref{figR}. We read off that $R \sim 0.2$ is preferred and the $\chi^2$
quickly becomes bad for larger $R$, mainly owing to $m_d$, $m_b$ and the
leptonic $\sin^2 \theta_{13}$. Also for very small $R$ the fit worsens, for
similar reasons as for large $R$, however, not in a dramatic way. On
the other hand, in the MSGUT there is a preferred range
$0.2 \lesssim R \lesssim 2$ and there is a genuine lower bound on $R$
as well~\cite{schwetz}.

With Eq.~(\ref{vud}) and the upper bound on $\sqrt{v_d^2 + v_u^2}$ in 
Eq.~(\ref{fit-normal}) we see that we are allowed to raise $w_R$ to 
$w_R \sim 6 \times M_\mathrm{GUT}$, without violating the
inequality~(\ref{test}). 

It has to be checked that our numerical solution given by 
Eq.~(\ref{solution-normal}) and Table~\ref{ratios} respects the
perturbative regime in the Yukawa sector. Since the procedure has been
explained in detail in~\cite{GK}, we confine ourselves to the
essentials. The two Higgs doublets of the MSSM, $H_d$ and $H_u$, have
hypercharges $+1/2$ and $-1/2$ and VEVs 
$v \cos \beta$ and $v \sin \beta$, respectively. The corresponding 
Yukawa coupling matrices are given by $Y_d = M_d/(v \cos \beta)$,
etc. It turns out that the largest Yukawa couplings are 
$\left( Y_u \right)_{33} \simeq \left( Y_D \right)_{33}$, 
where the largest contribution comes from $H'_{33}$. 
Using $\tan \beta = 10$, it
is given by $r_H H'_{33}/(v \sin \beta) \simeq -0.48$. This confirms
that the Yukawa couplings are safely in the perturbative regime.

Finally, we want to estimate the size of type~II seesaw contributions
to $\mnu$. The corresponding mass matrix is given by 
$w_L F'/v_d = w_L F'/(r_R M_\mathrm{GUT})$---see 
Eqs.~(\ref{onescale}), (\ref{mnu}) and (\ref{vud}). 
The largest element in $F'$ is the 33-element. With the value of this
element from Eq.~(\ref{solution-normal}) and $r_R$ from 
Table~\ref{ratios}, the type~II mass
matrix contributes at most
$\left( 1.2 \times 10^{-2} \right) \times w_L$.
Since we expect $w_L \sim v^2/M_\mathrm{GUT} \sim 10^{-3}$ eV, we find
the announced suppression with respect to type~I seesaw contributions.

\paragraph{A fit for the inverted neutrino mass spectrum:}
Searching for a fit by imposing the inverted ordering 
$m_3 < m_1 < m_2$ of the neutrino masses 
($\deltasol = m_2^2 - m_1^2$ as for the normal spectrum, but 
$\deltaatm = m_2^2 - m_3^2$), 
we find a solution which is even 
better as in the case of normal ordering. It has the following
properties: 
\begin{equation}\label{fit-inverted}
\begin{array}{l}
\chi^2 = 0.011, \quad
\sqrt{v_d^2 + v_u^2} = 8.46 \: \mbox{GeV}, \\[1mm]
m_1 = 4.920 \times 10^{-2} \: \mbox{eV}, \quad
m_2 = 5.000 \times 10^{-2} \: \mbox{eV}, \quad 
m_3 = 2.18 \times 10^{-4} \: \mbox{eV},
\end{array}
\end{equation}
with the matrices
\begin{eqnarray}
H' & = & \left( \begin{array}{ccc}
2.39744 & 0 & 0 \\ 
0 & 33.6387 & 0 \\ 
0 & 0 & 1127.980
\end{array} \right), \nonumber \\
G' & = & \left( \begin{array}{ccc}
0 & -2.41722 & -2.65793 \\ 
2.41722 & 0 & 0.0107775 \\ 
2.65793 & -0.0107775 & 0 
\end{array} \right),
\label{solution-inverted} \\
F' & = & \left( \begin{array}{ccc}
-1.05201 & -0.0960901 & 0.174940 \\ 
-0.0960901 & -14.0343 & 26.0245 \\ 
0.174940 & 26.0245 & -52.9848
\end{array} \right), \nonumber
\end{eqnarray}
where all numerical values are in units of MeV,
and the VEV ratios are displayed in Table~\ref{ratios}.
For all practical purposes the fit is perfect and there is no need to 
give any pull values.

Now we come to the predictions of our scenario in the case of the 
inverted neutrino mass spectrum. 
Concerning CP violation in neutrino oscillations, our best fit gives
$\delta_\mathrm{PMNS} = -107.6^\circ$. However, this value has no meaning
because $\chi^2$ as a function of $\delta_\mathrm{PMNS}$ is flat for all
practical purposes. The same is true for the leptonic quantities 
$\sin^2 \theta_{23}$ and $\sin^2 \theta_{13}$ in the physically relevant
ranges. However, there is a definite prediction for the neutrino mass
spectrum: hierarchy is strongly preferred---see Fig.~\ref{figR}. When the
quantity $R = m_3/\sqrt{\deltasol}$ becomes large the fit turns bad; however,
there is no clear-cut reason, it is mostly the down-quark masses and the 
top-quark mass which 
are not well reproduced and the fit value of the 
leptonic $\sin^2 \theta_{13}$ is around its experimental upper bound.

In the second relation of Eq.~(\ref{fit-inverted}) 
we have again used our reference value~(\ref{onescale}). 
Now inequality~(\ref{test}) is respected for 
$w_R \lesssim 20 \times M_\mathrm{GUT}$, i.e. there is more freedom
for $w_R$ than in the normal case.

As before, large Yukawa couplings in $Y_u$ and $Y_D$ are induced by
$H'_{33}$. But now, because $r_D$ is so big, a slightly larger
coupling is $\left( Y_D \right)_{13} 
\simeq i r_D G'_{13}/(v \sin \beta) \simeq i \times 0.66$, still in
the perturbative regime. The discussion of the smallness of type~II
seesaw contributions to $\mnu$ proceeds as for the normal spectrum.

In Eq.~(\ref{solution-inverted}) the elements $G'_{23}$, $F'_{12}$ and
$F'_{13}$ are rather small. This might suggest to set them zero, which
is achieved by the horizontal $\mathbbm{Z}_2$ symmetry 
$\psi_{L1} \to -\psi_{L1}$, $D_{pqr} \to - D_{pqr}$. However, this is
untenable because it would lead to vanishing $\delta_\mathrm{CKM}$.
The reason is that this horizontal $\mathbbm{Z}_2$ can be combined
with the CP transformation of Section~\ref{CPviolation} to a new
symmetry CP$^\prime$, under which the vacuum state of our scenario is
invariant.\footnote{Under CP$^\prime$, the VEVs $i \kappa_{d,u,\ell,D}$ do not
  change sign!} Consequently, in that case there is no CP
violation~\cite{branco,ecker} and one can show---as it must be in such
a case---that $\exp (i\delta_\mathrm{CKM}) = \pm 1$~\cite{ecker}. 

\section{Conclusions}
\label{concl}

In this paper we have investigated fermion masses and mixings in 
an SUSY $SO(10)$ scenario,\footnote{In our analysis SUSY enters only
  via the input parameters whose values we need at the GUT scale. In the
  evolution of these parameters from the electroweak to the GUT scale
  we assume the renormalization group equations of the MSSM.}
originally proposed in~\cite{dutta1}, where the Yukawa coupling 
matrices of the scalars in the irreps 
$\ten$, $\oht$ and $\hts$ are real and CP violation is
induced only by imaginary VEVs of the $\oht$. This gives a scenario
with 18 real parameters in the fermion mass sector. Recent results
from the MSGUT require a single heavy $SO(10)$ breaking scale, which
is then cogent for the MSGUT extended by the $\oht$ as well.

There are the following differences between Ref.~\cite{dutta1} and
the present paper:
Firstly, our $\mnu$ is induced by the type~I seesaw 
mechanism and type~II is negligible,
whereas in~\cite{dutta1} it was assumed that type~II dominates. 
Secondly, we use a purely numerical method, employing the minimization
of a $\chi^2$ function, whereas Ref.~\cite{dutta1} uses an approximate 
semianalytical method. 

We have found excellent fits to fermion masses and mixings for both types
of neutrino mass spectra. We want to emphasize this in particular for
the inverted mass spectrum, for which the system of fermion mass 
matrices in the MSGUT---which has no $\oht$---does 
not allow an acceptable fit~\cite{schwetz}, 
though complex Yukawa couplings and VEVs and 
contributions to $\mnu$ from both seesaw types are admitted.\footnote{The
  MSGUT system has 13 absolute values and 8 phases.}
The fits presented in this paper have the following 
features. The diagonal Yukawa coupling matrix $H$ of the $\ten$ is strongly 
hierarchical and is responsible, in the charged-fermion mass spectra,
for the hierarchy between 2nd and 3rd families.
The correct size of the neutrino masses is reproduced by
a cooperation of two effects: rather large contributions $r_H H + ir_D G$
to the neutrino Dirac-mass matrix $M_D$ 
from the couplings of the $\ten$ and $\oht$, 
where $G$ is the Yukawa coupling matrix of the $\oht$,
and a moderately small coupling matrix $F$ of the
$\hts$, which enters with its inverse in the type~I seesaw formula. 
The contribution of the $\oht$ to the charged-fermion masses
and to the CKM matrix is rather small, whereas $r_D G$ in $M_D$ introduces
large leptonic mixing angles.
Similar features were found in the previous sample fit of~\cite{GK},
though the assumptions concerning the fermion mass matrices in that
paper are quite different from those in the present paper, 
apart from the use of spontaneous CP violation in both scenarios.

Unfortunately, our scenario is not very predictive. However, it does have one
clear-cut prediction, namely hierarchy for both the
normal and inverted neutrino mass spectrum. This is quantified by the
observable $R$ in Fig.~\ref{figR} from where we read off 
$m_\mathrm{min} \ll \sqrt{\deltasol}$.

Apparently, extending the MSGUT by the $\oht$ leads to an ambiguous
situation concerning fermion mass matrices: quite different
assumptions can result in excellent fits. Whether these fits are compatible
with the NMSGUT~\cite{newMSGUT}, where the VEVs are subject to certain
relations, remains to be checked. One aspect seems to
emerge: spontaneous CP violation plays in important role in
both the fermion mass matrices~\cite{GK} and the spontaneous
breaking~\cite{newMSGUT} of $SO(10)$.

\vspace{5mm}

\noindent
\textbf{Acknowledgment:} W.G. is grateful to L.~Lavoura for useful 
discussions.

\newpage

\appendix
\setcounter{equation}{0}
\renewcommand{\theequation}{A\arabic{equation}}

\section{The spinor representation of $SO(10)$}
\label{so10}

A possible representation---on the space $\mathcal{H}$ 
of the fivefold tensor product of
$\mathbbm{C}^2$---for the Clifford algebra associated with
the Lie algebra $so(10)$ is given by~\cite{wilczek}
\begin{equation}\label{gamma}
\Gamma_{2j-1} = 
\sigma_3^{(5-j)} \otimes \sigma_1 \otimes \mathbbm{1}^{(j-1)},
\quad
\Gamma_{2j} = 
\sigma_3^{(5-j)} \otimes \sigma_2 \otimes \mathbbm{1}^{(j-1)}
\quad (j=1,\ldots,5),
\end{equation}
where a superscript $(k)$ denotes the $k$-fold tensor product.
The $2 \times 2$ matrices $\sigma_i$ ($i=1,2,3$) are the Pauli
matrices and $\mathbbm{1}$ denotes the $2 \times 2$ unit matrix.
It is easy to check that the $\Gamma_p$ ($p = 1, \ldots, 10$) fulfill
\begin{equation}\label{gammaprop}
\{ \Gamma_p,\Gamma_q\} = 2 \delta_{pq} \mathbbm{1}^{(5)} 
\quad \mbox{and} \quad
\quad \Gamma_p = \Gamma_p^\dagger = \Gamma_p^{-1}.
\end{equation}
It is well known that the matrices
\begin{equation}\label{sigma}
{\textstyle \frac{1}{2}} \sigma_{pq} \equiv 
{\textstyle \frac{1}{2}} \Gamma_p \Gamma_q
\quad 1 \leq p < q \leq 10,
\end{equation}
have precisely the same commutation relations as the basis elements 
\begin{equation}\label{soN}
(M_{pq})_{jk} = \delta_{pj} \delta_{qk} - \delta_{qj} \delta_{pk},
\quad 1 \leq p < q \leq N,
\end{equation}
of $so(10)$. The $\sigma_{pq}$ generate the spinor irrep $\mathbf{16}$
of $so(10)$ on the 16-dimensional space 
\begin{equation}
\frac{1}{2} \left( \mathbbm{1}^{(5)} + \Gamma_{11} \right) \mathcal{H}
\quad \mbox{with} \quad  
\Gamma_{11} = \sigma_3^{(5)}.
\end{equation}
Note that $\Gamma_{11}$ anticommutes with all $\Gamma_p$ ($p = 1,
\ldots, 10$).

For the Yukawa couplings one needs the matrix $\B$ and its following
properties: 
\begin{equation}\label{B}
\mathcal{B} \equiv \Gamma_1 \Gamma_3 \Gamma_5 \Gamma_7 \Gamma_9, 
\quad
\B = \B^\dagger = \B^{-1}, \quad
\Gamma_p \B = \xi_p \B \Gamma_p 
\quad \mbox{with} \quad \xi = (-1)^{1+p}.
\end{equation}

\end{document}